\documentclass[prb,aps,twocolumn,showpacs,superscriptaddress]{revtex4}  

\usepackage{graphicx}

\usepackage[dvipdf,
            colorlinks,
            hyperindex]
            {hyperref}

\begin{document}

\title{Spin unrestricted linear scaling electronic structure theory\\
  and its application to magnetic carbon doped BN nanotubes}

\author{H. J. Xiang}
\affiliation{Hefei National Laboratory for Physical Sciences at
  Microscale, 
  University of Science and Technology of
  China, Hefei, Anhui 230026, People's Republic of China}

\affiliation{USTC Shanghai Institute for Advanced Studies,
  University of Science and Technology of China,
  Shanghai 201315, People's Republic of China}

\author{W. Z. Liang}
\affiliation{Hefei National Laboratory for Physical Sciences at
  Microscale, 
  University of Science and Technology of
  China, Hefei, Anhui 230026, People's Republic of China}

\author{Jinlong Yang}
\thanks{Corresponding author. E-mail: jlyang@ustc.edu.cn}

\affiliation{Hefei National Laboratory for Physical Sciences at
  Microscale, 
  University of Science and Technology of
  China, Hefei, Anhui 230026, People's Republic of China}

\affiliation{USTC Shanghai Institute for Advanced Studies,
  University of Science and Technology of China,
  Shanghai 201315, People's Republic of China}

\author{J. G. Hou}
\affiliation{Hefei National Laboratory for Physical Sciences at
  Microscale, 
  University of Science and Technology of
  China, Hefei, Anhui 230026, People's Republic of China}

\author{Qingshi Zhu}
\affiliation{Hefei National Laboratory for Physical Sciences at
  Microscale, 
  University of Science and Technology of
  China, Hefei, Anhui 230026, People's Republic of China}
\affiliation{USTC Shanghai Institute for Advanced Studies,
  University of Science and Technology of China,
  Shanghai 201315, People's Republic of China}

\date{\today}

\begin{abstract}
We present an extension of density matrix based linear scaling
electronic structure theory to incorporate spin degrees of
freedom. When the spin multiplicity of the system can be predetermined,
the generalization of the existing linear scaling methods
to spin unrestricted cases is straightforward.
However, without calculations it is hard to
determine the spin multiplicity of some complex systems,
such as, many magnetic nanostuctures,
some inorganic or bioinorganic molecules.
Here we give a general prescription to obtain the
spin-unrestricted ground state of open shell systems.
Our methods are implemented into the linear scaling
trace-correcting density matrix purification algorithm.
The numerical atomic orbital basis, rather than the commonly adopted
Gaussian basis functions is used.       
The test systems include O$_2$ molecule,
and magnetic carbon doped BN(5,5) and BN(7,6) nanotubes.
Using the newly developed method, we find the magnetic moments in
carbon doped BN nanotubes couple antiferromagnetically with each
other. Our results suggest that the linear scaling spin-unrestricted
trace-correcting purification method is very powerful to treat large
magnetic systems.
\end{abstract}

%\pacs{71.15.-m, 31.15.Ne, 75.75.+a, 36.40.Cg}
%31.15.Ne Self-consistent-field methods
%36.40.Cg Electronic and magnetic properties of clusters
%71.15.-m Methods of electronic structure calculations
%75.75.+a Magnetic properties of nanostructures
\maketitle

\section{Introduction}

Recently, the spin of the electron has caused a resurgence
of interest because it promises a wide variety of new devices 
that combine logic, storage and sensor applications. 
Moreover, these ``spintronic''
devices might lead to quantum computers and quantum communication
based on electronic solid-state devices.\cite{spintronics}
To explain the experimental findings and predict novel magnetic
properties of nanostructures, {\it ab initio} electronic structure
calculations on these magnetic systems are indispensable.
However, {\it ab initio} electronic structure
calculations are usually limited to small and medium size 
molecular systems. The obstacle lies in rapidly increasing 
computational costs as the systems become larger and more complex.
Usually, the molecules which possess novel magnetic
properties include few hundreds or thousands of atoms.
They are too large to be
calculated by conventional {\it ab initio} or density functional theory.
For instance, single-molecule magnets (SMMs), \cite{SMM1} which
have attracted much interest for the quantum tunneling of
magnetization in such systems, are typically very large molecules,
e.g., [Mn$_{12}$O$_{12}$(O$_2$CC$_6$F$_5$)$_16$(H$_2$O)$_4$]$^-$,
\cite{SMM2} a SMM of Mn$_{12}$ family, is composed by $260$ atoms.
Extensively studies on such systems clearly necessitate new methods
with desired computational complexity.

Linear scaling (O(N)) electronic structure theory in combination
with tight-binding, self-consistent Hartree-Fock (HF) or density functional
theory (DFT) has become a very powerful tool to investigate
very large complex systems,\cite{ref1,ref2} such as,
the silicon systems,\cite{ref20} the giant fullerenes,
the low-dimension nanomaterials,\cite{ref21} and the biological
molecules.\cite{ref22}
They have been successfully applied to calculate the molecular energies,
obtain the optimized molecular geometries and evaluate the static and
dynamical molecular properties of very large systems.\cite{ref1}
Many linear scaling methods have already been developed to
 build the effective Hamiltonian\cite{ON_H}
and to avoid cubic scaling
Roothaan step by replacing it with diagonalization-free alternatives.
\cite{ref3,ref4,ref5,ref6,ref7,ref8,ref9,ref10,ref11,ref12,ref13,ref14,ref15,ref16,ref17,ref18}
After the algorithms whose computational cost scales asymptotically
only linearly with the size of the system are available
for building the effective Hamiltonian,                      
the Roothaan step which updates the occupied spaces becomes 
the rate-determining step in large enough self-consistent-field
(SCF) calculations.
Typically this may occur for calculations involving several thousand
basis functions. Two of the main alternatives may be briefly
summarized as follows. 
One may attempt to update the one-particle
density matrix
itself,\cite{ref3,ref4,ref5,ref9,ref10,ref11,ref12,ref13,ref14,ref15,ref16,ref17,ref18} rather than
the molecular orbitals. 
Recent examples include 
the improved Fermi Operator expansion method\cite{ref13,ref14,ref15}
and the purification projection
schemes which was first proposed by Palser and Manolopoulos\cite{ref12} 
and later
was improved by Niklasson and his coworkers.\cite{ref16,ref17,ref18}
Second, one may attempt to obtain
localized molecular orbitals,\cite{ref6,ref7,ref8} rather than the
delocalized orbitals that diagonalize the Hamiltonian matrix.
Some systematic comparisons of different approaches have been
reported in the context of semiempirical electronic structure methods.
\cite{ON_semi}

However, to the best of our knowledge, linear scaling
methods are all exclusively applied to closed shell systems.
As the first attempt of our efforts in the field,
we provide a first survey on the possibility
of applying the linear scaling methods to open shell systems 
at self-consistent electronic structure level.
We notice that if the spin multiplicity of the
insulating open shell systems can be predetermined, almost all linear
scaling methods are readily applicable. 
But sometimes, without calculations it is hard to
determine the spin multiplicity of some complex systems,
such as, some inorganic or bioinorganic molecules
with the transition metal atoms involved.  
Usually one needs to calculate few states assuming
different spin multiplicities. By comparing
their energies one finally determine the spin degree of the systems.
It is inconvenient and time-consuming.
It therefore extremely necessary to develop
a method which can automatically determine 
spin multiplicity of large complex systems while
has low computational complexity.

In this work, we develop a general method to determine automatically the spin
multiplicity in the 
calculation and extend the Niklasson's 
trace-correcting density matrix purification algorithm (TC2)
to deal with spin unrestricted systems. Here, we select the TC2
purification algorithm to demonstrate and implement our new spin
unrestricted linear scaling methods since the TC2 purification is very
simple and efficient at both low ($< 10\%$) and high ($> 90\%$)
occupancies. Interestingly, an intriguing density matrix perturbation
theory\cite{DMPT1} based on the TC2 purification is proposed and
succefully applied to calculate electric polarizability recently.\cite{DMPT2}
Our methods are implemented in a DFT program employing localized numerical
atomic orbitals as basis sets.\cite{ref28,siesta_manual}
This paper is organized as follows: 
In the next section, 
we shall describe the spin 
unrestricted linear scaling electronic structure theory.
We illustrate our method by presenting
a spin unrestricted version of the TC2 algorithm.
In Sec.~\ref{result}, we describe the details of the
implementation and perform some test calculations to illustrate the
rightness, robustness, and linear scaling behavior of our methods.
Magnetic carbon doped 
BN nanotubes are studied with the spin unrestricted TC2 method.
Our results clearly indicate that the magnetic moments in
carbon doped BN nanotubes couple antiferromagnetically with each
other.
We discuss the possibility of applying the spin unrestricted linear
scaling methods to magnetic metallic systems in Sec.~\ref{dis}.
Finally, our concluding remarks are given in Sec.~\ref{conclusion}.

\section{Theory}
The most general one-particle reduced density matrix may be written as  
\begin{equation}
  \begin{array}{ll}
    \rho(\mathbf{x},\mathbf{x'})= & \rho_{\alpha
      \alpha}(\mathbf{r},\mathbf{r'})+ \rho_{\alpha
      \beta}(\mathbf{r},\mathbf{r'})+ \\
    & \rho_{\beta \alpha}(\mathbf{r},\mathbf{r'})+
    \rho_{\beta \beta}(\mathbf{r},\mathbf{r'}),
  \end{array}
\end{equation}
here $\mathbf{x}$ ($\mathbf{x'}$) is a combination of a space 
coordinate $\mathbf{r}$ ($\mathbf{r'}$)
and a spin coordinate $s$ ($s'$). It is shown by McWeeny \cite{ref23} 
that in any state
where the $z$ component of total spin is definite, the two components 
$\rho_{\alpha \beta}(\mathbf{r},\mathbf{r'})$ and 
$\rho_{\beta \alpha}(\mathbf{r},\mathbf{r'})$ must vanish, and the
first-order reduced density matrix has the form:
\begin{equation}
  \rho(\mathbf{x},\mathbf{x'})=
  \rho_{\alpha}(\mathbf{r},\mathbf{r'})+
  \rho_{\beta}(\mathbf{r},\mathbf{r'}), 
\end{equation}
for simplicity, we write $\rho_{\alpha \alpha}(\mathbf{r},\mathbf{r'})$ 
as $\rho_{\alpha}(\mathbf{r},\mathbf{r'})$,
and define the similar abbreviation  for 
$\rho_{\beta \beta}(\mathbf{r},\mathbf{r'})$. 
In case where the number
of electrons for both spin components are integers,
\begin{equation}
  \begin{array}{ll}
    \rho_{\alpha}^{2}  =  \rho_{\alpha} & ~~ Tr(\rho_{\alpha})=N_{\alpha}
    \\
    \rho_{\beta}^{2}  =  \rho_{\beta}  & ~~ Tr(\rho_{\beta})=N_{\beta} \\
    N_e=N_{\alpha}+N_{\beta}, &
  \end{array}
\end{equation}
here $N_{\alpha}$ and $N_{\beta}$ are the number
of electrons for spin up and spin down components respectively, and
$N_{e}$ represents the total number of electrons.  
It indicates that if $N_{\alpha}$ and $N_{\beta}$ are known prior, 
one can deal with open shell systems without any problem using the
linear scaling methods applied to close shell systems.
For example, given $N_{\alpha}$ and $N_{\beta}$, we can deal
with open shell systems using LNVD density
matrix minimization (DMM),\cite{ref4,ref5}
simplified density matrix minimization (SDMM),\cite{ref11} or spectral 
projection approach. 
As an example, here we give a spin unrestricted TC2 algorithm with
predetermined spin 
multiplicity (PSUTC2) to show how
it works: 
\begin{equation} 
\rho_{\alpha,n+1}(\rho_{\alpha,n}) =
\left\{\begin{array}{ll}
\rho_{\alpha,n}  ^2, &  Tr(\rho_{\alpha,n} ) \geq N_{\alpha} \\
2\rho_{\alpha,n} - \rho_{\alpha,n}^2, & Tr(\rho_{\alpha,n}) < N_{\alpha}
\end{array} \right.
\end{equation}
with $\rho_{\alpha,0}  =(\epsilon_{N}(H_{\alpha}) I - H _{\alpha} ) / 
(\epsilon_{\alpha,0} - \epsilon_{0}(H_{\alpha}) )$ ($H_{\alpha}$ is the
majority part of the Hamiltonian matrix)
where the constants  $\epsilon_{0}(H_{\alpha})$ and $\epsilon_{N}(H_{\alpha})$ are the
lowest and highest eigenvalues of $H_\alpha$, respectively.
The purification algorithm  for $\rho_{\beta}$ is similar to the above
step just with $\alpha$ replaced by $\beta$.
We note that McWeeny has extended his density matrix method to deal
with open shell systems with given orbital occupations,\cite{ref23} 
however not in the framework of linear scaling methods.
However, sometimes the above algorithm is inconvenient in practical
applications since it needs the predefined spin multiplicity. One
shortcoming is 
that to find the electronic ground states, one must carry out several
calculations with different spin multiplicity. Another serious problem
is that one may encounter fractional occupation in practical
calculations if the spin multiplicity is not the right one with no
fractional occupation. Under this circumstance, 
the algorithm might not be linear scaling using real space
representations, or even fail since in this case the density matrix is
not idempotent. 

Here we give a prescription to solve this problem.
We define new $H$ and $\rho$ as:
\begin{equation}
  H =
\left( \begin{array}{cc}
H_{\alpha} & 0  \\
0 & H_{\beta} 
\end{array} \right) \qquad 
\rho =
\left( \begin{array}{cc}
\rho_{\alpha} & 0  \\
0 & \rho_{\beta}
\end{array} \right).
\end{equation}
Here the new operators satisfy $H \rho =\rho H$, $\rho \rho = \rho$,
and $Tr(\rho)=N_e$.
Through the newly defined Hamiltonian $H$, the spin multiplicity 
for the ground state can be determined automatically according to the
'Aufbau' principle.
It is easy to see that if we find $\rho$ which minimizes $Tr(H
\rho)$ within the above constraints, $\rho_{\alpha}$ and
$\rho_{\beta}$ can be extracted easily from $\rho$.
In fact, the minimization problem can be 
solved by many linear scaling methods for closed shell systems.
One difference is that we now deal with matrices with the dimension
$2N_b$, here $N_b$ is the number of basis sets.
However, due to the block form of $H$ and $\rho$, the $2N_b$ dimension
problem can be decomposed into two $N_b$ dimension problems.
Here we detail this method by presenting an algorithm
based on TC2 projection algorithm.
The spin unrestricted TC2 (SUTC2)
projection algorithm is given by this pseudocode:
\begin{equation}
\begin{array}{l}
{\it subroutine~ } {\it SUTC2}( H_{\alpha},H_{\beta}, \rho_{\alpha}, 
\rho_{\beta},N_e, {\it ErrorLimit})\\
{\it estimate~} \varepsilon_0(H_{\alpha}),~\varepsilon_N(H_{\alpha}), 
\varepsilon_0(H_{\beta}),~\varepsilon_N(H_{\beta}) \\
\varepsilon_N=max(\varepsilon_N(H_{\alpha}),~\varepsilon_N(H_{\beta}))\\
\varepsilon_0=min(\varepsilon_0(H_{\alpha}),~\varepsilon_0(H_{\beta})) \\
\rho_{\alpha,0} = (\varepsilon_N I-H_{\alpha})/(\varepsilon_N-\varepsilon_0) \\
\rho_{\beta,0} = (\varepsilon_N I-H_{\beta})/(\varepsilon_N-\varepsilon_0) \\
{\it while~ Error > ErrorLimit} \\ 
~~ {\it if}~ Tr [\rho_{\alpha,n}+\rho_{\beta,n} ] - N_e < 0 \\
~~ ~~ \rho_{\alpha,n+1} = 2\rho_{\alpha,n} - \rho_{\alpha,n}^2    \\
~~ ~~ \rho_{\beta,n+1} = 2\rho_{\beta,n} - \rho_{\beta,n}^2  \\
~~ {\it else}\\
~~ ~~ \rho_{\alpha,n+1} = \rho_{\alpha,n}^2 \\
~~ ~~ \rho_{\beta,n+1} =  \rho_{\beta,n}^2 \\
~~ {\it end}\\
~~ {\it estimate~ Error} \\
{\it end}\\
\rho_{\alpha} = \rho_{\alpha,n}  \\
\rho_{\beta} = \rho_{\beta,n} ~ . \\
\end{array}
\end{equation}
The scheme can be described as follows: First using the same
scaling factors we normalize $H_{\alpha}$ and $H_{\beta}$ to get
initial matrices $\rho_{\alpha,0}$ and $\rho_{\beta,0}$ with all theirs
eigenvalues in the range $[0,1]$. Then $\rho_{\alpha,n}$ and
$\rho_{\beta,n}$ are updated in the same way depending on the sum of
the traces. In this way, $\rho_{\alpha}$ and $\rho_{\beta}$ is
obtained using the same purification polynomial. The monotonicity of
the purification polynomial leads to the correct occupations according
to the 'Aufbau' principle. 

Although we illustrate our method by presenting only the SUTC2 method,
however, our method is quite general and can be easily generalized to
many other density matrix or localized orbitals based linear scaling 
methods. For instance, the KMG functional \cite{ref7} can be easily
adapted to include 
spin degrees of freedom given the chemical potential of the magnetic
insulating systems.
 
\section{Implementation and Results}
\label{result}
\subsection{Implementation}
Both schemes developed in this work for dealing with open shell
systems are implemented in SIESTA, a standard Kohn-Sham
density functional program using norm-conserving pseudopotentials
and numerical atomic orbitals as basis sets.\cite{ref28}
In SIESTA, periodic boundary conditions are employed to simulate
both isolated and periodic systems.
Here since we aim at large systems, $\Gamma$-point sampling is used.
There is a linear scaling solver using localized, Wannier-like
orbitals employing the KMG functional\cite{ref7} in SIESTA. Unfortunately, the
convergence of the 
conjugate gradient (CG) minimization of the electronic energy in the first
SCF step might be extremely slow (up to 2000 CG iterations, compared to 20 
in further SCF steps). 
Another inconvenience is the chemical potential must be given prior to
conserve the total charge, which might be notably difficult for small gap
systems.\cite{siesta_manual} Thus we implement the robust density
matrix purification 
methods in SIESTA.
Saravanan {\it et al.} \cite{ref26} showed that the multiatom blocked
sparse matrix 
multiplications can be much faster than a standard element-by-element
sparse matrix package and also more efficient than the atom-blocked sparse
matrix algebra.\cite{ref11}   
The blocking scheme benefits from the use of highly-optimized
level-3 basic linear algebra subroutines (BLAS) for large
submatrix multiplications.  
So in this work, we employ the blocked compressed sparse row
(BCSR)\cite{ref11,ref26,cpc} data 
structure with multiatom blocks for sparse matrix computations.
We use a multiatom blocked sparse matrix multiplications
with dropping (filtering) of multiatom
blocks with the Frobenious norm below a numerical
threshold($10^{-4}-10^{-6}$) to obtain energy as accurate as 1
mHatree.
Relative to the cutoff approach, a major advantage of
threshold metered sparse linear algebra is that it avoids
discontinuities in the potential energy surface associated with
atoms moving in and out of the cutoff radius.
We work in an orthogonal representation though in siesta $H$ is 
evaluated in the nonorthogonal basis of atomic orbitals (AO).
We achieve this by transforming the AO Hamiltonian matrix $H_{AO}$
to an orthonormal basis using $H=ZH_{AO}Z^{T}$ and obtaining the AO
density matrix $\rho_{AO}$ using $\rho_{AO}=Z^{T} \rho Z$, where the
inverse factor $Z=L^{-1}$, and $L$ is the Cholesky factor for which
$S=LL^{T}$. The inverse factor $Z$ is obtained directly using the
state of the art blocked approximate inverse (AINV)
algorithm.\cite{ref27,BAINV}  
As for the force and stress calculations, only the orthogonality parts
one must care about in our density matrix implementation.
The orthogonality force and stress require the energy-density matrix  
$E=\rho_{AO}H_{AO}S^{-1}$. In our implementation, the energy-density
matrix is calculated as $E=((\rho_{AO}H_{AO})Z^{T})Z$ and it is only
calculated when the SCF reaches its
convergence.  

All calculations reported here employ the local density approximation
(LDA).\cite{LDA} And no structural optimizations are performed in all
following calculations to save computing resource.  
First we validate our methods and our implementation by calculate 
oxygen molecule with the well known triplet ground state. As expected,
by specifying the spin triplet state, using the PSUTC2 method, we can
get the same energy for the triplet state as that from the
diagonalization calculation.  
And the SUTC2 method can also give the triplet ground state with the same
energy for oxygen molecule without given the electronic occupation.   

\subsection{Performance of the method}
\begin{figure}
  \includegraphics[width=8cm]{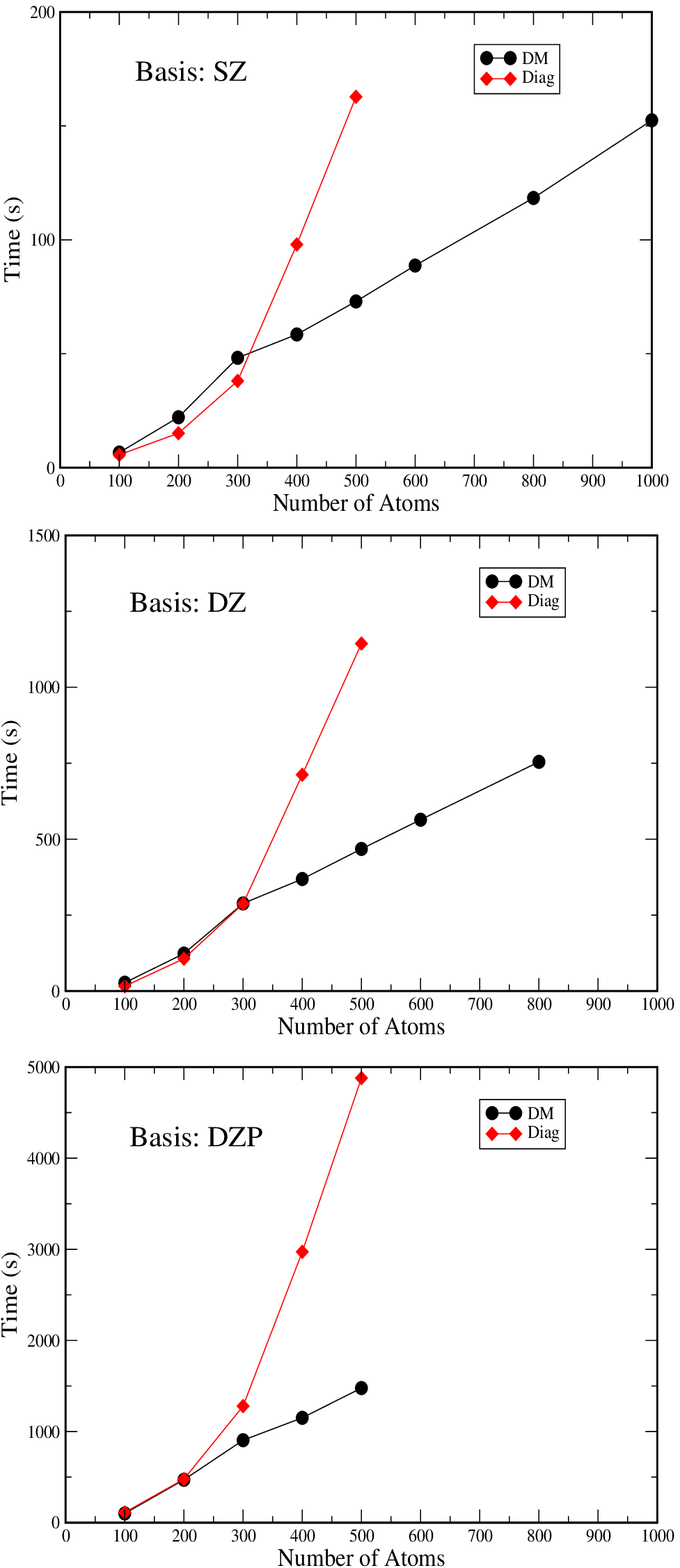}
  \caption{CPU time per SCF cycle for carbon doped
    BN(5,5) nanotube 
    using the PSUTC2 method and
    traditional diagonalization method with different basis sets.
    All calculations were carried out on a 1.5 GHz 
    Itanium 2 CPU processor running RedHat Linux Advanced Server V2.1.}
  \label{fig1}
\end{figure}

A recent DFT calculation indicates that
carbon doping induces spontaneous magnetization in BN nanotubes.\cite{c_BN}
The density of states (DOS) for both carbon substituted BN(5,5) nanotube
and carbon substituted BN(9,0) nanotube display the insulating or
semiconducting behavior. So carbon doping BN nanotubes serve as 
ideal systems to test our methods. We choose BN(5,5)
nanotubes with a boron atom substituted by a carbon for this purpose.
Both PSUTC2 and SUTC2 methods are tested.
In the PSUTC2 method, the magnetic moment of these systems is fixed to
1 $\mu_{B}$.\cite{c_BN}
For these systems, we find that the PSUTC2 method is faster by 25\%
than the SUTC2 method. In TC2 methods, the efficiency is determined by the
magnitude of the band gap.\cite{ref16}
In the PSUTC2 method, the convergence for the spin up density
matrix is determined by the band gap of the majority part, and so does
for the spin down component. However, the convergence in the SUTC2
method depends on the magnitude of the system band gap, which is
always smaller in ferromagnetic (FM) systems. For antiferromagnetic (AFM)
systems, the system band gap are 
the same as the band gaps for both spin components, and the SUTC2
method should be as efficient as the PSUTC2 method.  
So, the PSUTC2 method is prefered for systems where the spin
multiplicity is known prior.
For systems without knowledge for the spin multiplicity, one should
use the SUTC2 method to account for the magnetic moment of the systems.  
Furthermore, if we assume the spin multiplicity is not changed in all
SCF cycles, we can combine the efficiency of the PSUTC2 method and the
robustness of the SUTC2 method by using SUTC2 method in the first SCF
cycle and using PSUTC2 method in the following SCF
cycles. Fig.~\ref{fig1} shows the CPU time per SCF cycle for the
diagonalization and PSUTC2 methods. Three different basis sets
(single-$\zeta$ (SZ),
double-$\zeta$ (DZ), double-$\zeta$ plus polarization functions (DZP))
are employed. The diagonalization method clearly shows a 
O(N$^3$) scaling, in contrast sharply to the linear scaling behavior
of the PSUTC2 method for all basis sets. The critical system size 
where the PSUTC2 method is faster than the diagonalization method is
dependent on the basis set employed. For instance, the critical system
size is 320, 300, and 200 for SZ, DZ and DZP basis sets respectively. 

\subsection{Applications to magnetic carbon doped BN(5,5) and BN(7,6)
  nanotubes} 
Following the recent discovery of ferromagnetism at
room temperature in an all-carbon system consisting of polymerized
C60,\cite{ref29} there has been increased interest in magnetism
in metal-free systems. Although previous experiment 
indicated a preference for zig-zag and near zig-zag BN tubes,\cite{zig-zag}      
a very recent high-resolution electron diffraction study on BN nanotubes
grown in a carbon-free chemical vapor deposition process
revealed a dispersion of the chiral angles $\alpha$ 
in the ranges of $10^{\circ} \le \alpha \le 15^\circ$ and
$25^{\circ} \le \alpha \le 30^\circ$.\cite{BN_chiral} Chiral BN
nanotubes usually 
contain large number of atoms in a unit cell, e.g., BN(7,6) nanotube
have 508 atoms in a unit cell with chiral angle
$\alpha=27.46^\circ$. These nanotubes are difficult to be 
treated using traditional methods. Here we conduct a linear scaling
spin unrestricted calculation on BN(7,6) nanotube with a boron atom
substituted by a carbon. The DZ basis set is employed for this
purpose. For better comparison with carbon doped BN(5,5) nanotube, we
show the structure and  calculated 
spin density for both carbon doped BN(5,5) nanotube with 500 atoms per cell
and and carbon doped BN(7,6) nanotube. Clearly, the distribution of
spin density for both carbon doped nanotubes is basically the same:
The spin density is mainly contributed by carbon 2$p_z$ orbital and 
the nitrogen atoms near carbon have some small magnetic
moments. 
To learn more knowledge about the electronic and magnetic properties of
the system, we obtain the density of states (DOS) by a non-SCF
calculation using diagonalization. 
Fig.~\ref{fig3} shows the total DOS (TDOS) as well as carbon partial
DOS (PDOS) for both carbon doped BN(5,5) nanotube with 500 atoms per
cell and carbon doped BN(7,6) nanotube. We can see that
in the plotting energy range, in both cases carbon has 
sizable contribution only to the two states around the Fermi level. 
The shape of the DOS is very similar for both carbon doped BN
nanotubes. 
Also these DOS resemble that of carbon doped BN(5,5) nanotube with 80
atoms per cell
provided by Wu {\it et al.} \cite{c_BN} except that the band gap in
our case is smaller possibly due to the LDA functional we employed.

Previous calculations indicated there exist magnetic moments in carbon
doped BN nanotubes and our calculations confirm it. Then, how the
local magnetic moments in carbon doped BN nanotubes couple with each other:
ferromagnetically or antiferromagnetically?
Here we conduct some calculations on BN(7,6) nanotube with two
boron atoms substituted by two carbon atoms to address this issue.
Six different configurations (A, B, C, D, E, and F) are considered, as shown in
Fig.~\ref{fig4}(a). 
The total energy of both AFM and FM states
for these six carbon doped BN nanotubes is shown in
Fig.~\ref{fig4}(b).
Generally, the relatively energy difference converges much faster
along with the cutoff radius for the density matrix than 
the absolute total energy. In fact, some test calculations indicate
that the deviation of the energy difference from the exact value is
smaller than 10 meV. 
As shown in Fig.~\ref{fig4}(b), structure B with AFM state is the most stable
configuration among these six carbon doped BN(7,6) nanotubes.  
Moreover, we can see that when two carbon atoms are in the same hexagonal
ring, the AFM state is more favorable over the FM state. In addition,
the energy difference for B is larger than that for A possibly due to  
the fact that the two carbon 2$p_z$ orbitals in B are more parallel
and thus the superexchange AFM interaction in B is larger. 
If the distance between two carbon atoms is larger than 3 \AA, i.e.,
two carbon atoms are not in the same hexagonal ring, the coupling
between two magnetic moments is negligibly small. 
So from Fig.~\ref{fig4}(b), we can conclude that 
the magnetic moments in carbon doped BN nanotubes couple
antiferromagnetically with each other.     

\section{Discussion}
\label{dis}
Recently, developing linear scaling methods to deal with metallic
systems has attracted much interest.
Baer and Head-Gordon developed an energy renormalization-group method,
with which the computational effort scales near linearly with system
size even when the density matrix is highly nonlocal.\cite{renorm}
Goedecker {\it et al.} showed that in addition to the real-space
localization of the density matrix, there is also a localization in
Fourier space. \cite{wavelet} 
Within the multiresolution wavelet representation it was shown how the
sparsity of the density matrix is preserved for localized insulating
systems as well as for itinerant metallic systems. 
\cite{wavelet}
Combining such techniques with our methods, O(N) calculations on
metallic magnetic systems might be possible since the number of matrix
multiplications in the TC2 purification, essentially is independent of
system size even for metallic systems.\cite{ref16} 
Although, there are no unfilled shells in spin unrestricted
HF theory, as proved by Bach {\it et al.}, \cite{SUHF}
however, for spin unrestricted DFT methods, sometimes fractional occupation is
possible. 
In principle, the fractional occupation problem can be treated
using PM\cite{ref12} or TRS4\cite{ref18} purification. So the
generalized spin unrestricted 
form of PM or TRS4 purification can treat these systems.

\section{Conclusions}
\label{conclusion}
To conclude, we give a first survey on applying linear scaling
electronic structure 
methods to spin unrestricted systems.
Two methods are proposed to deal with systems with or without
predetermined spin multiplicity respectively.
We demonstrate our methods by detailing the PSUTC2 and SUTC2
projection algorithms. 
The current methods have been implemented in a Kohn-Sham
density functional program employing numerical atomic orbitals as
basis sets. We apply our method to deal with magnetic carbon
doped BN nanotubes. Carbon doped BN(7,6) nantoube has similar magnetic
properties as carbon doped BN(5,5) nantoube. 
Moreover, FM coupling is unfavorable for carbon doped BN nantoubes.
The results suggest that our methods pave the way
for carrying out linear scaling calculations on spin unrestricted
systems, such as magnetic nanostructures.

\section*{ACKNOWLEDGEMENTS}
This work is partially supported by the National Project for the
Development of Key Fundamental Sciences in China (G1999075305), by
the National Natural Science Foundation of China
(50121202, 10474087), by the USTC-HP HPC project, by
the EDF of USTC-SIAS, and by the SCCAS.

\clearpage

\begin{figure}
 \includegraphics[width=8cm]{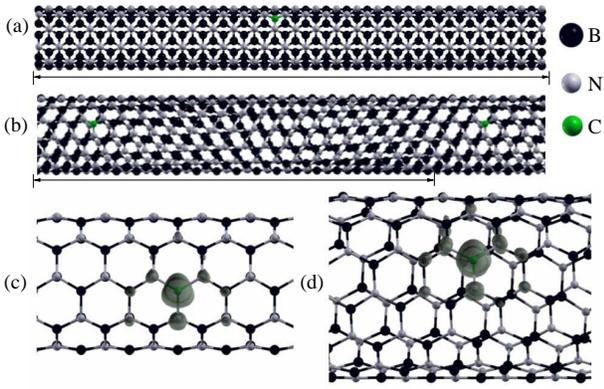}
  \caption{Structures for (a) carbon doped BN(5,5) nanotube
    with 500 atoms per cell and (b) carbon
    doped BN(7,6) nanotube. The regions between the vertical lines 
    indicate the unit cells. 
    Spin densities of the two nanotubes are shown in (c) and (d) respectively.}
  \label{fig2}
\end{figure}
%\clearpage
\begin{figure}
  \includegraphics[width=8cm]{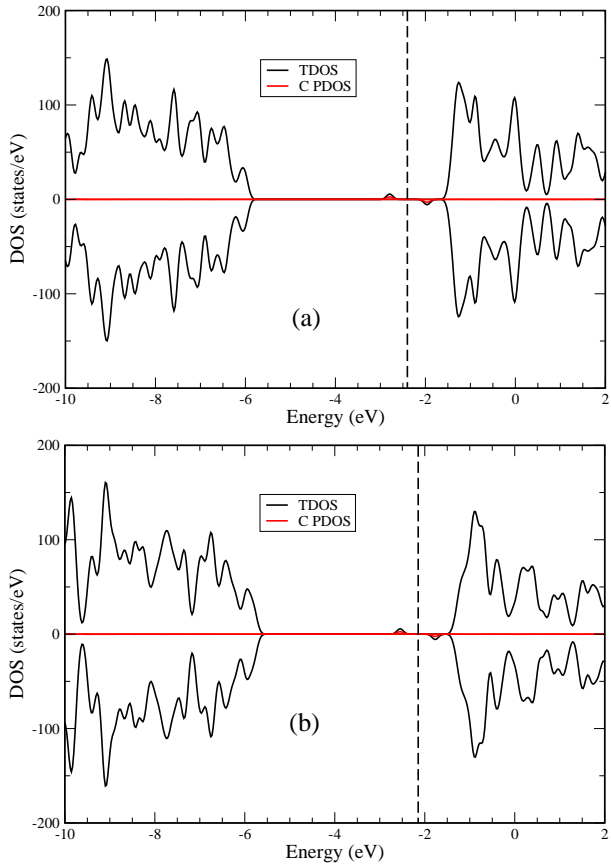}
  \caption{ 
    Majority and minority spin DOS of (a)
    carbon doped BN(5,5) nanotube with 500 atoms per cell and (b) carbon
    doped BN(7,6) nanotube. TDOS and carbon PDOS are shown in 
    black and red solid lines respectively.
    The Fermi energy is indicated by
    the vertical dashed line.}
  \label{fig3}
\end{figure}

%\clearpage

\begin{figure}
  \includegraphics[width=8cm]{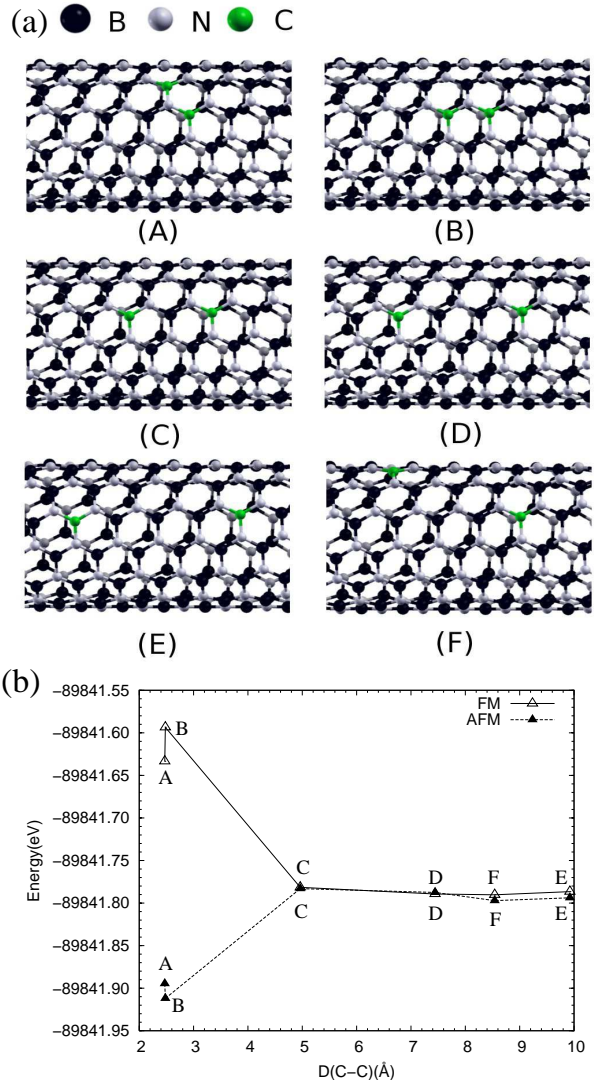}
  \caption{(a) Six different structures A, B, C, D, E,
    and F for BN(7,6) tubes
    with two boron atoms substituted by two carbon atoms. (b) shows
    the total energy for both AFM and FM states for different
    carbon doped BN(7,6) tubes.
    The horizontal axis is the nearest distance between two carbon
    atoms. } 
  \label{fig4}
\end{figure}

\end{document}